\documentclass[11pt,superscriptaddress,aps,prd,preprint,showpacs]{revtex4}
\usepackage[dvips]{graphicx}
\usepackage{amsfonts}
\usepackage{slashed}

\begin{document}

\title{Dynamical Lorentz symmetry breaking in a tensor bumblebee model}

\author{J. F. Assun\c c\~ao}
\author{T. Mariz}
\affiliation{Instituto de F\'\i sica, Universidade Federal de Alagoas,\\ 57072-900, Macei\'o, Alagoas, Brazil}
\email{jfassuncao,tmariz@fis.ufal.br}

\author{J. R. Nascimento}
\author{A. Yu. Petrov}
\affiliation{Departamento de F\'{\i}sica, Universidade Federal da Para\'{\i}ba,\\
 Caixa Postal 5008, 58051-970, Jo\~ao Pessoa, Para\'{\i}ba, Brazil}
\email{jroberto,petrov@fisica.ufpb.br}


\begin{abstract}
In this paper, we formulate a theory of the second-rank antisymmetric (pseudo)tensor field minimally coupled to a spinor, calculate the one-loop effective potential of the (pseudo)tensor field, and, explicitly, demonstrate that it is positively defined and possesses a continuous set of minima, both for tensor and pseudotensor cases.  Therefore, our model turns out to display the dynamical Lorentz symmetry breaking. We also argue that, contrarily to the derivative coupling we use here, derivative-free couplings of the antisymmetric tensor field to a spinor do not generate the positively defined potential and thus do not allow for the dynamical Lorentz symmetry breaking.
\end{abstract}


\maketitle

\section{Introduction}

As it is well known, the Lorentz symmetry breaking can be introduced in three manners, the explicit one, where the constant vector or tensor introducing privileged spacetime direction is added from the very beginning, the anomalous one, where the spacetime possesses nontrivial topology allowing for a natural arising of Lorentz-breaking terms, and the spontaneous one, where the constant vector or tensor emerges as a vacuum expectation of some vector or tensor field, respectively. While the first manner became paradigmatic, being used to formulate the Lorentz-breaking extension of the standard model~\cite{Colladay:1996iz,Colladay:1998fq}, and the second one allowed for a new, very interesting mechanism of arising the Carroll-Field-Jackiw (CFJ) term, essentially involving the nonperturbative methodology \cite{Klinkhamer:1999zh,Klinkhamer:2017hms}, the interest to the third, spontaneous manner, is based on the fact that this approach provides a mechanism allowing to explain the origin of Lorentz symmetry breaking. Namely,  this way was originally proposed in~\cite{Kostelecky:1989jp} (see also~\cite{Kostelecky:1989jw}), where the Lorentz symmetry breaking was introduced for the first time being suggested to arise in the low-energy limit of string theory. 

The first vector field theory model, involving a potential allowing for spontaneous Lorentz symmetry breaking, was introduced in~\cite{Kostelecky:2000mm}. In~\cite{Bertolami:2005bh}, where this model was denominated as the bumblebee model for the first time, it was generalized to curved spacetime, and some solutions of modified Einstein equations in the presence of spontaneous Lorentz symmetry breaking were obtained. Further, various issues related to the vector bumblebee model, including the case of curved background, were considered, see f.e.~\cite{Altschul:2005mu,Seifert:2009gi}, and in \cite{Seifert:2009gi} it has also been argued that spontaneous Lorentz symmetry breaking in curved space is the most appropriate way to introduce Lorentz-violating extension of gravity. 

The aspects of dynamical Lorentz symmetry breaking, occurring due to perturbative corrections, have been treated in~\cite{Gomes:2007mq,Assuncao:2017tnz}. In \cite{Gomes:2007mq}, by performing the fermion integration of a self-interacting massive vector theory, the vector bumblebee model  with the Lagrangian
\begin{eqnarray}\label{BB0}
{\cal L}_B &=& -\frac{1}{12}F_{\mu\nu} F^{\mu\nu} -\frac{\lambda}{4}\left(B_{\mu} B^{\mu}-\beta^2\right)^2,
\end{eqnarray}
was obtained, however, with $\lambda<0$ (where $F_{\mu\nu}=\partial_\mu B_\nu-\partial_\nu B_\mu$, and $\beta^2=\beta_\mu\beta^\mu$). This situation was overcome in~\cite{Assuncao:2017tnz}, where a massless theory and the exact propagator allowing to take into account all orders of the expansion in the constant $\beta_\mu$ were considered so that the $\lambda$ was shown to be positive, i.e., the potential is positively defined in this latter approach.

Thus, the spontaneous Lorentz symmetry breaking has been relatively well studied for vector field models. Therefore, taking into account that, namely due to this mechanism, constant tensors of various ranks which break the Lorentz symmetry, can arise  \cite{KosSam}, it is interesting to investigate the spontaneous Lorentz symmetry breaking for more generic tensor field models. Although a systematic approach to this study has been proposed already in~\cite{Altschul:2009ae}, up to now there are very few results obtained for higher-rank Lorentz-breaking tensor field models, with mostly tree-level aspects being considered, see f.e.~\cite{Colatto:2003he,Hernaski:2016dyk,Aashish:2018aqn,Aashish2}.

Therefore, it is natural to generalize the methodology developed for the vector bumblebee model, for the analogous theory of the antisymmetric tensor field, which can display the spontaneous Lorentz symmetry breaking as well (for different issues related to the antisymmetric tensor field, without context of the Lorentz symmetry breaking, see \cite{Kalb:1974yc,Freedman:1980us}, and references therein). Originally the bumblebee model on the base of the antisymmetric tensor field theory was introduced in \cite{Altschul:2009ae}. Here we present its simplest version with the quartic potential looking like:
\begin{eqnarray}\label{BB}
{\cal L}_B &=& -\frac{1}{12}H_{\mu\nu\lambda} H^{\mu\nu\lambda} + \bar\psi(i\slashed{\partial}-ieB^{\mu\nu}\gamma_{[\mu}\partial_{\nu]}\gamma_5^q-m)\psi-\frac{\lambda}{4}\left(B_{\mu\nu} B^{\mu\nu}-\beta^2\right)^2,
\end{eqnarray}
where $H_{\mu\nu\lambda}=\partial_\mu B_{\nu\lambda} + \partial_\nu B_{\lambda\mu}+\partial_{\lambda}B_{\mu\nu}$ is a stress tensor for $B_{\mu\nu}$, $\gamma_{[\mu}\partial_{\nu]}=\frac12(\gamma_{\mu}\partial_{\nu}-\gamma_{\nu}\partial_{\mu})$, and $q=1,2$, with $\gamma_5^2=1$. We note that this coupling differs from the spinor-tensor interactions considered in~\cite{HariDass:2001dp,Pilling:2002ij} and~\cite{Diamantini:2013yka,Kim:2018obm}, where the vertices look like $i\bar{\psi}\epsilon_{\mu\nu\lambda\rho}H^{\mu\nu\lambda}\gamma^{\rho}\gamma_5\psi$ and $\bar{\psi}\epsilon_{\mu\nu\lambda\rho}H^{\mu\nu\lambda}\gamma^{\rho}\psi$, respectively, involving the stress tensor $H_{\mu\nu\lambda}$ rather than the $B_{\mu\nu}$ itself. So, unlike these couplings, our interaction can be treated as a minimal one. 

We observe that only our coupling $i\bar\psi B^{\mu\nu}\gamma_{[\mu}\partial_{\nu]}\gamma_5^q\psi$ allows us for obtaining a potential for $B_{\mu\nu}$, while other ones yield contributions depending on stress tensor only, which justifies our choice namely of this coupling for the study of the dynamical Lorentz symmetry breaking. Then, we introduce the spontaneous Lorentz symmetry breaking in the standard way, that is, we shift  the bumblebee field $B_{\mu\nu}$  by the rule $B_{\mu\nu} \to \beta_{\mu\nu}+B_{\mu\nu}$, where  $\langle B_{\mu\nu}\rangle = \beta_{\mu\nu}$ is a non-trivial vacuum expectation value (VEV) of $B_{\mu\nu}$, and $\beta^2=\beta_{\mu\nu}\beta^{\mu\nu}$. So, the Lagrangian (\ref{BB}) becomes
\begin{eqnarray}
\label{ourmodel}
{\cal L}_B &=& -\frac{1}{12}H_{\mu\nu\lambda}H^{\mu\nu\lambda} + \bar\psi(i\slashed{\partial}-ib^{\mu\nu}\gamma_{[\mu}\partial_{\nu]}\gamma_5^q-ieB^{\mu\nu}\gamma_{[\mu}\partial_{\nu]}\gamma_5^q-m)\psi \nonumber\\
&&-\frac{\lambda}{4}\left(B_{\mu\nu}B^{\mu\nu}+\frac{2}{e}B_{\mu\nu} b^{\mu\nu}\right)^2,
\end{eqnarray}
where $b_{\mu\nu}=e\beta_{\mu\nu}$. Thus, we see that the spontaneous Lorentz violation in (\ref{BB}) implied arising of the new term $i\bar\psi b^{\mu\nu}\gamma_{[\mu}\partial_{\nu]}\gamma_5^q\psi$. This term is nothing more as a particular form of the Lorentz-breaking extension of free spinor action, introduced in \cite{Colladay:1998fq}, for the case when $b^{\mu\nu}$ (denoted there as $d^{\mu\nu}$ and $c^{\mu\nu}$, for $q=1$ and $q=2$, respectively) is antisymmetric. Although in most studies this coefficient is assumed to be symmetric, see f.e. \cite{Carroll:2008pk}, there is no reason forbidding it to be antisymmetric.
We note that the situation with $q=2$, i.e., when $b^{\mu\nu}$ is a constant tensor and not a pseudotensor, allows for a possibility to remove the coupling of the spinor to the antisymmetric  $b_{\mu\nu}$ term in the case of a free spinor theory \cite{Coll2002}, through some transformation of the spinor field. However, in the case of a nontrivial interaction between a spinor and a dynamical antisymmetric field, this transformation will generate additional spinor-tensor vertices which clearly modify quantum contributions. Hence, even at $q=2$ the quantum impact of this new term is nontrivial, and, certainly, it will be the case for $q=1$, where the new term $i\bar\psi b^{\mu\nu}\gamma_{[\mu}\partial_{\nu]}\gamma_5\psi$ cannot be ruled out.

The structure of the paper looks like follows. In section 2, the effective potential is calculated for tensor and pseudotensor cases, and the possibility of having minima is discussed. In section 3, we obtain the kinetic term and the bumblebee potential for the (pseudo)tensor field. Finally, section 4 is a Summary where our results are discussed.

\section{Effective potential and its minima}

As we already said, different issues related to the bumblebee model have been studied in a number of papers (besides of the works cited above, see also, e.g., Refs.~\cite{Maluf:2014dpa,Nascimento:2014vva,Hernaski:2014jsa,Maluf:2015hda,Escobar:2017fdi}). In this work, we will follow the idea originally proposed in~\cite{Coleman:1973jx}, that the quantum corrections can give origin to the spontaneous symmetry breaking, and will show that the bumblebee potential for the tensor field can be dynamically induced through radiative corrections from a self-interacting fermion theory, given by the Lagrangian
\begin{eqnarray}\label{Lag}
{\cal L}_0 &=& \bar\psi (i\slashed{\partial}-m)\psi - \frac{G}{2}J_{\mu\nu}J^{\mu\nu},
\end{eqnarray}
where the current is $J_{\mu\nu}=i\bar\psi\gamma_{[\mu}\partial_{\nu]}\gamma_5^q\psi$, as follows from (\ref{ourmodel}). Indeed, it is convenient to introduce an auxiliary field $B_{\mu\nu}$, in order to eliminate the term $J_{\mu\nu}J^{\mu\nu}$, with $G=\frac{e^2}{g^2}$, so that the above expression can be rewritten as
\begin{eqnarray}
\label{rewrite}
{\cal L} &=& {\cal L}_0 + \frac{g^2}{2} \left(B_{\mu\nu}-\frac{e}{g^2}J_{\mu\nu}\right)^2 \nonumber \\
&=& \frac{g^2}{2}B_{\mu\nu} B^{\mu\nu} + \bar\psi(i\slashed{\partial}-ieB^{\mu\nu}\gamma_{[\mu}\partial_{\nu]}\gamma_5^q-m)\psi.
\end{eqnarray}

In this section, we generate the bumblebee potential in a very simple way. In order to obtain the effective action, and consequently the bumblebee effective potential, we start with the generating functional
\begin{eqnarray}
Z(\bar \eta,\,\eta) &=& \int DB_\mu D\psi D\bar\psi e^{i\int d^4x({\cal L}+\bar\eta\psi+\bar\psi\eta)}\nonumber\\
&=& \int DB_{\mu\nu} e^{i\int d^4x \frac{g^2}{2}B_{\mu\nu}B^{\mu\nu}}\int D\psi D\bar\psi e^{i\int d^4x(\bar\psi S^{-1}\psi+\bar\eta\psi+\bar\psi\eta)},
\end{eqnarray}
where $S^{-1}=i\slashed{\partial}-ieB^{\mu\nu}\gamma_{[\mu}\partial_{\nu]}\gamma_5^q-m$, is the operator describing the quadratic action.  Now, by performing the shift of the fermionic fields, $\psi\rightarrow \psi-S\eta$ and $\bar{\psi}\rightarrow \bar{\psi}-\bar{\eta}S$, so that $\bar\psi S^{-1}\psi+\bar\eta\psi+\bar\psi\eta \rightarrow \bar\psi S^{-1}\psi-\bar\eta S \eta$, we obtain
\begin{eqnarray}
Z(\bar \eta,\,\eta) &=& \int DB_{\mu} e^{i\int d^4x \frac{g^2}{2}B_{\mu\nu}B^{\mu\nu}}\int D\psi D\bar\psi e^{i\int d^4x(\bar\psi S^{-1}\psi-\bar\eta S \eta)}.
\end{eqnarray}
Finally, integrating over fermions, we get
\begin{equation}\label{GF}
Z(\bar \eta,\,\eta) = \int DB_\mu \exp\left(iS_\mathrm{eff}[B] - i\int d^4x\, \bar\eta\, S\, \eta \right),
\end{equation}
where the effective action is given by
\begin{equation}\label{Seff}
S_\mathrm{eff}[B] = \frac{g^2}{2} \int d^4x\, B_{\mu\nu} B^{\mu\nu} -i \mathrm{Tr} \ln(\slashed{p}-eB_{\mu\nu}\gamma^{[\mu}p^{\nu]}\gamma_5^q-m).
\end{equation}
The $\mathrm{Tr}$ symbol stands for the trace over Dirac matrices as well as for integrating over momentum or coordinate spaces. The matrix trace can be readily calculated, so that for the effective potential, we have
\begin{equation}\label{Vef}
V_\mathrm{eff} = -\frac{g^2}{2}B_{\mu\nu} B^{\mu\nu} +i \,\mathrm{tr} \int\frac{d^4p}{(2\pi)^4}\, \ln(\slashed{p}-eB_{\mu\nu}\gamma^{[\mu}p^{\nu]}\gamma_5^q-m).
\end{equation}
We note that, unlike the case of the vector field \cite{Assuncao:2017tnz}, here, we have no essential simplifications in the massless case, since there is no convenient exact form for the massless spinor propagator in the case of its dependence on the constant second-rank tensor.

The nontrivial minima of this potential can be obtained as usual, from the condition of vanishing the first derivative of the potential: 
\begin{equation}\label{DVef}
\frac{dV_\mathrm{eff}}{dB_{\mu\nu}}\Big|_{B_{\mu\nu}=\beta_{\mu\nu}} =  -\frac{g^2}{e} b^{\mu\nu} - ie\,\Pi^{\mu\nu} = 0,
\end{equation}
where, again, $b_{\mu\nu}=e \beta_{\mu\nu}$ and the one-loop tadpole amplitude is 
\begin{eqnarray}\label{Tad}
\Pi^{\mu\nu} &=& \mathrm{tr} \int\frac{d^4p}{(2\pi)^4} \frac{1}{\slashed{p}-b_{\alpha\beta}\gamma^{[\alpha}p^{\beta]}\gamma_5^q-m} \gamma^{[\mu} p^{\nu]}\gamma_5^q.
\end{eqnarray}

 Let us calculate the above expression by expanding the propagator in terms of $b_{\alpha\beta}$ and considering, initially, $q=1$. This situation is more involved than for $q=2$, since we must use dimensional regularization together with 't Hooft and Veltman prescription \cite{tHooft:1972tcz}. To do this, we first extend the 4-dimensional spacetime to a $D$-dimensional one, so that $d^4p/(2\pi)^4$ goes to $\mu^{4-D}d^Dp/(2\pi)^D$, where $\mu$ is an arbitrary scale parameter with the mass dimension 1. In the following, we introduce the anticommutation relation $\{\gamma^{\mu},\gamma^{\nu}\}=2g^{\mu\nu}$, with the contraction $g_{\mu\nu} g^{\mu\nu}=D$. Then, we split the $D$-dimensional Dirac matrices $\gamma^{\mu}$ and the $D$-dimensional metric tensor $g^{\mu\nu}$ into 4-dimensional parts and $(D-4)$-dimensional parts, i.e., $\gamma^{\mu}=\bar{\gamma}^{\mu}+\hat{\gamma}^{\mu}$ and $g^{\mu\nu}=\bar{g}^{\mu\nu}+\hat{g}^{\mu\nu}$, and consider the commutation relations $\{\bar{\gamma}^{\mu},\gamma^{5}\}=0$ and $[\hat{\gamma}^{\mu},\gamma^{5}]=0$. Thus, using the identity
\begin{eqnarray}
\frac{\slashed{p}+m}{p^2-m^2} b_{\alpha\beta}\gamma^{[\alpha}p^{\beta]}\gamma_5 \frac{\slashed{p}+m}{p^2-m^2}  &=& \frac{(\slashed{p}+m)(\bar\slashed{p}-\hat\slashed{p}+m)}{(p^2-m^2)^2} b_{\alpha\beta}\gamma^{[\alpha}p^{\beta]}\gamma_5 \nonumber\\
&=& \frac{(\slashed{p}+m)(\slashed{p}-2\hat\slashed{p}+m)}{(p^2-m^2)^2} b_{\alpha\beta}\gamma^{[\alpha}p^{\beta]}\gamma_5,
\end{eqnarray} 
where $\hat\slashed{p}=p_\kappa \gamma_\lambda \hat g^{\kappa\lambda}$, with $g_{\kappa\lambda}\hat g^{\kappa\lambda}=D-4$, we can write the tadpole amplitude (\ref{Tad}) as a series in $b_{\alpha\beta}$, i.e., $\Pi_{\mu\nu} = \sum\limits_n \Pi_{\mu\nu}^{(2n+1)}$, where
\begin{eqnarray}\label{Pi2n+15}
\Pi_{\mu\nu}^{(2n+1)} &=& \mathrm{tr}\int\frac{d^4p}{(2\pi)^4} \frac{[(\slashed{p}+m)(\slashed{p}-2\hat\slashed{p}+m)]^{n+1}}{(p^2-m^2)^{2n+2}}(b_{\alpha\beta}\gamma^{[\alpha} p^{\beta]}\gamma_5)^{2n+1} \,\gamma_{[\mu}p_{\nu]}\gamma_5.
\end{eqnarray}
Note that only odd contributions of $b_{\alpha\beta}$ survive, since in the opposite case we have $\mathrm{tr}[(\slashed{p}+m)(\slashed{p}-2\hat\slashed{p}+m)]^n(b_{\alpha\beta}\gamma^{[\alpha} p^{\beta]}\gamma_5)^{2n}(\slashed{p}+m)\gamma_{[\mu}p_{\nu]}\gamma_5 = (-1)^n(b_{\alpha\beta}p^\beta)^{2n}\mathrm{tr}[(\slashed{p}+m)(\slashed{p}-2\hat\slashed{p}+m)]^n(\slashed{p}+m)\gamma_{[\mu}p_{\nu]}\gamma_5$, which vanishes for any $n$.

To study the minima of effective potential, in the usual case of weak field, so that the potential can be presented as a power series in the field, it is sufficient to consider only two lower contributions to $\Pi_{\mu\nu}$, that is, those ones with $n=0$ and $n=1$, which are linear and cubic in the field $b^{\mu\nu}$, respectively.
So, for $n=0$, we have
\begin{eqnarray}
(-ie)\Pi_{\mu\nu}^{(1)} = -\frac{5e m^4}{8\pi^2} \left(\frac{1}{\epsilon'}+\frac{3}{4}\right) b_{\mu\nu},
\end{eqnarray}
whereas for $n=1$, we obtain
\begin{eqnarray}
(-ie)\Pi_{\mu\nu}^{(3)} = \frac{35e m^4}{48\pi^2} \left(\frac{1}{\epsilon'}+\frac{3}{4} \right) (b_{\mu\nu} b_{\alpha\beta}b^{\alpha\beta}+2b_{\mu\alpha}b_{\nu\beta}b^{\alpha\beta}),
\end{eqnarray}
where $\frac 1{\epsilon'}=\frac 1\epsilon-\ln\frac m{\mu'}$, with $\epsilon=4-D$ and $\mu'^2=4\pi\mu^2e^{-\gamma}$. This last contribution can be simplified by using the expression 
\begin{equation}\label{b3}
b_{\mu\alpha}b_{\nu\beta}b^{\alpha\beta} = \frac12 b_{\mu\nu}b_{\alpha\beta}b^{\alpha\beta} +\frac14 \tilde{b}_{\mu\nu}b_{\alpha\beta}\tilde{b}^{\alpha\beta},
\end{equation}
where $\tilde{b}^{\mu\nu}=\frac12 \epsilon^{\mu\nu\kappa\lambda}b_{\kappa\lambda}$.

Then, the gap equation (\ref{DVef}) can be rewritten as
\begin{equation}\label{DVef25}
\frac{dV_\mathrm{eff}}{dB_{\mu\nu}}\Big|_{B_{\mu\nu}=\beta_{\mu\nu}} =  \left[-\frac{e}{G} -\frac{5e m^4}{8\pi^2}\left(\frac{1}{\epsilon'}+\frac{3}{4} \right)\left(1-\frac{7}{3}x_1\right)\right]b^{\mu\nu} +\frac{35e m^4}{96\pi^2}\left(\frac{1}{\epsilon'}+\frac{3}{4} \right)x_2\, \tilde{b}^{\mu\nu} +\cdots = 0,
\end{equation}
where $x_1=b_{\alpha\beta}b^{\alpha\beta}$ and $x_2=b_{\alpha\beta}\tilde{b}^{\alpha\beta}$. 

Let us now calculate the tadpole amplitude (\ref{Tad}) for $q=2$. To do this, we use the identity 
\begin{eqnarray}
\frac{\slashed{p}+m}{p^2-m^2} b_{\alpha\beta}\gamma^{[\alpha}p^{\beta]} \frac{\slashed{p}+m}{p^2-m^2}  &=& \frac{(\slashed{p}+m)(-\slashed{p}+m)}{(p^2-m^2)^2} b_{\alpha\beta}\gamma^{[\alpha}p^{\beta]} \nonumber\\
&=& -\frac{b_{\alpha\beta}\gamma^{[\alpha}p^{\beta]}}{p^2-m^2},
\end{eqnarray} 
so that we can also write the tadpole amplitude as a series in $b_{\alpha\beta}$, i.e., $\Pi_{\mu\nu} = \sum\limits_n \Pi_{\mu\nu}^{(2n+1)}$, where
\begin{eqnarray}\label{Pi2n+1}
\Pi_{\mu\nu}^{(2n+1)} = \mathrm{tr}\int\frac{d^4p}{(2\pi)^4} \frac{(-1)^{n+1}}{(p^2-m^2)^{n+1}} (b_{\alpha\beta}\gamma^{[\alpha}p^{\beta]})^{2n+1}\,\gamma_{[\mu}p_{\nu]}.
\end{eqnarray}
 Again, only odd contributions of $b_{\alpha\beta}$ survive because in the opposite case the result is proportional to $\mathrm{tr}(b_{\alpha\beta}\gamma^{[\alpha}p^{\beta]})^{2n}(\slashed{p}+m)\gamma_{[\mu}p_{\nu]}=(b_{\alpha\beta}p^\beta)^{2n}\,\mathrm{tr}(\slashed{p}+m)\gamma_{[\mu}p_{\nu]}$, which vanishes for any $n$.

Now, writing $(b_{\alpha\beta}\gamma^{[\alpha}p^{\beta]})^{2n+1} = (b_{\rho\sigma}p^\sigma)^{2n}\,b_{\alpha\beta}\gamma^{[\alpha}p^{\beta]}$, we can easily calculate the trace in (\ref{Pi2n+1}). Then, we obtain
\begin{equation}
\Pi_{\mu\nu}^{(2n+1)} = 4 \int\frac{d^4p}{(2\pi)^4} \frac{(-1)^{n+1}}{(p^2-m^2)^{n+1}}  (b_{\rho\sigma}p^\sigma)^{2n}\, b_{[\mu\beta}p^\beta p_{\nu]},
\end{equation}
where $b_{[\mu\beta}p^\beta p_{\nu]}=\frac12(b_{\mu\beta}p^\beta p_{\nu}-b_{\nu\beta}p^\beta p_{\mu})$ is the product antisymmetrized with respect to $\mu$ and $\nu$. In order to calculate the integral, we use the Feynman formula 
\begin{eqnarray}\label{int}
\int\frac{d^Dp}{(2\pi)^D} \frac{p_{\mu_1} \cdots p_{\mu_p}}{(p^2-m^2)^\alpha} = \frac{i(-1)^{-\frac D2}}{(4\pi)^{\frac{D}{2}}} \frac{\Gamma(\alpha-\frac{D}{2}-\frac{p}{2})}{2^{\frac p2}\Gamma(\alpha)} (-m^2)^{\frac{D}{2}+\frac{p}{2}-\alpha} \sum_{\mathrm{perm}}g_{\mu_1\mu_2}g_{\mu_3\mu_4} \cdots g_{\mu_{p-1}\mu_p},
\end{eqnarray}
where the sum is taken over all permutations, with $\alpha=n+1$ and $p=2n+2$, so that we get
\begin{eqnarray}\label{Pi2n+1:2}
\Pi_{\mu\nu}^{(2n+1)} &=& \frac{4i\mu^{4-D}(-1)^{n+1}(-1)^{-\frac D2}}{(4\pi)^{\frac{D}{2}}2^{n+1}\Gamma(n+1)} \Gamma\left(-\frac{D}{2}\right)(-m^2)^{\frac{D}{2}}\ b_{\rho_1\mu_1}{b^{\rho_1}}_{\mu_2} \cdots  b_{\rho_n\mu_{2n-1}}{b^{\rho_n}}_{\mu_{2n}} \nonumber\\
&&\times b_{[\mu\mu_{2n+1}} \sum_{\mathrm{perm}}g^{\mu_1\mu_2}g^{\mu_3\mu_4} \cdots {g^{\mu_{2n+1}}}_{\nu]}.
\end{eqnarray}
We note that in this case we can easily obtain all orders of expansion, while in the $q=1$ case, we can only calculate the amplitude order by order (see Eq.~(\ref{Pi2n+15})), but the complete sum apparently cannot be found in a closed form.

Thus, taking into account (\ref{Pi2n+1:2}), for $n=0$ and $n=1$, we have
\begin{eqnarray}
(-ie)\Pi_{\mu\nu}^{(1)} = -\frac{e m^4}{8\pi^2} \left(\frac{1}{\epsilon'}+\frac{3}{4} \right) b_{\mu\nu},
\end{eqnarray}
and
\begin{eqnarray}
(-ie)\Pi_{\mu\nu}^{(3)} = \frac{e m^4}{16\pi^2} \left(\frac{1}{\epsilon'}+\frac{3}{4} \right) (b_{\mu\nu} b_{\alpha\beta}b^{\alpha\beta}+2b_{\mu\alpha}b_{\nu\beta}b^{\alpha\beta}),
\end{eqnarray}
so that the gap equation (\ref{DVef}) takes the form
\begin{equation}\label{DVef2}
\frac{dV_\mathrm{eff}}{dB_{\mu\nu}}\Big|_{B_{\mu\nu}=\beta_{\mu\nu}} =  \left[-\frac{e}{G} -\frac{em^4}{8\pi^2}\left(\frac{1}{\epsilon'}+\frac{3}{4} \right)(1-x_1)\right]b^{\mu\nu} +\frac{em^4}{32\pi^2}\left(\frac{1}{\epsilon'}+\frac{3}{4} \right)x_2\, \tilde{b}^{\mu\nu} +\cdots = 0.
\end{equation}

Let us discuss the implications of the above equations (\ref{DVef25}) and (\ref{DVef2}). First of all, we note that when we integrate them, we arrive at the result looking like a linear combination of $(B_{\mu\nu}B^{\mu\nu})^2$ and $(B_{\mu\nu}\tilde{B}^{\mu\nu})^2$, with both coefficients accompanying these terms being positive. Hence, our effective potentials are positive definite and thus display minima (the explicit form of one of these potentials will be given further). Second, while the first term possesses a minimum at some definite value of $x_1$ different from zero, which allows to find the square of $b_{\mu\nu}$, the second term displays the minimum at $x_2=0$, which imposes an additional restriction on relations between components of the $b_{\mu\nu}$. At the same time, it is interesting to note that if we take the second derivative of $V_\mathrm{eff}$ in both cases, we find that its lower order in $b_{\mu\nu}$ will be proportional to $\eta_{\mu\nu}$, with the positive sign, i.e., we indeed have a minimum. It means that different values of $b_{\mu\nu}$, either in tensor or pseudotensor cases, will correspond to different vacua, and hence the spontaneous Lorentz symmetry breaking is possible in both situations.

It is worth mentioning that if the current is derivative-free, $J_{\mu\nu}=i\bar{\psi}\sigma_{\mu\nu}\gamma_5^q\psi$, instead of $J_{\mu\nu}=i\bar\psi\gamma_{[\mu}\partial_{\nu]}\gamma_5^q\psi$ we are considering in this paper, the result for (\ref{DVef}) is
\begin{equation}\label{DVef2J2}
\frac{dV_\mathrm{eff}}{dB_{\mu\nu}}\Big|_{B_{\mu\nu}=\beta_{\mu\nu}} =  \left[-\frac{e}{G} +\frac{em^2}{\pi^2}\left(\frac{1}{\epsilon'}-\frac{1}{2} \right) -\frac{2ex_1}{3\pi^2\epsilon'}\right]b^{\mu\nu} -\frac{2ex_2}{3\pi^2\epsilon'}\tilde{b}^{\mu\nu} +\cdots = 0,
\end{equation}
for both possibilities, $q=1$ and $q=2$. In this expression, we can observe that the terms proportional to $x_1b_{\mu\nu}$ and $x_2\tilde{b}_{\mu\nu}$ (or, after integrating the equation (\ref{DVef2J2}), to $(B_{\mu\nu}B^{\mu\nu})^2$ and $(B_{\mu\nu}\tilde{B}^{\mu\nu})^2$, respectively) have a negative sign, which indicates that the potential is not positive definite, and, hence, it does not display minima. This perception will be clearer below, when we will integrate the gap equation~(\ref{DVef2}) to obtain the potential.

In order to get more information about the Eq.~(\ref{DVef2}), let us try to evaluate the Eq.~(\ref{Tad}), for $q=2$, in a general way, by writing it as
\begin{equation}\label{Tad2}
\Pi^{\mu\nu} = \mathrm{tr} \int\frac{d^4p}{(2\pi)^4} \frac{\slashed{p}'+m}{p'^2-m^2} \gamma^{[\mu} p^{\nu]} = 2\int\frac{d^4p}{(2\pi)^4} \frac{p'^\mu p^\nu -p^\mu p'^\nu}{p'^2-m^2},
\end{equation} 
where $p'_\alpha=M_{\alpha\beta}p^\beta$, with $M_{\alpha\beta}=g_{\alpha\beta}-b_{\alpha\beta}$. Thus, we have $d^4p'=\mathrm{det}\left(\frac{\partial p'^\mu}{\partial p^\nu}\right)d^4p$, i.e., $d^4p'=\mathrm{det}\left(M^{\mu\alpha}g_{\alpha\nu}\right)d^4p=-\mathrm{det}\left(M^{\mu\alpha}\right)d^4p$, so that 
\begin{equation}
d^4p = -\mathrm{det}^{-1}\left(M^{\mu\alpha}\right) d^4p'.
\end{equation}
Now, as $p_\alpha=(M^{-1})_{\alpha\beta}p'^\beta$, we must calculate $(M^{-1})_{\alpha\beta}$, which is given by
\begin{equation}\label{Mm1}
(M^{-1})_{\alpha\beta} = \left[\left(1+\frac{x_1}2\right)g_{\alpha\beta} +b_{\alpha\beta}+b_{\alpha\gamma}{b^\gamma}_\beta-\frac{x_2}4\tilde{b}_{\alpha\beta}\right]\left(1+\frac{x_1}2-\frac{x_2^2}{16}\right)^{-1},
\end{equation}
where we have used the expression (\ref{b3}). Then, the Eq.~(\ref{Tad2}) can be rewritten as
\begin{equation}
\Pi^{\mu\nu} = -\mathrm{det}^{-1}(M^{\kappa\lambda})\left(g^{\mu\beta}(M^{-1})^{\nu\alpha}-(M^{-1})^{\mu\alpha}g^{\nu\beta}\right) \int\frac{d^4p'}{(2\pi)^4} \frac{p'_\alpha p'_\beta}{p'^2-m^2}.
\end{equation}
Finally, by using the expression (\ref{int}) and (\ref{Mm1}), as well as the fact that $\mathrm{det}(M^{\kappa\lambda}) = -\left(1+\frac{x_1}2\right)$, we obtain the gap equation
\begin{eqnarray}\label{DVef3}
\frac{dV_\mathrm{eff}}{dB_{\mu\nu}}\Big|_{B_{\mu\nu}=\beta_{\mu\nu}} = -\frac{e}{G}b^{\mu\nu} -\frac{e m^4}{8\pi^2} \left(\frac{1}{\epsilon'}+\frac{3}{4} \right) \left(1+\frac{x_1}2\right)^{-1}\left(1+\frac{x_1}2-\frac{x_2^2}{16}\right)^{-1}\left(b^{\mu\nu}-\frac{x_2}4\tilde{b}^{\mu\nu}\right) = 0. \;\;
\end{eqnarray}
We can observe that this Eq.~(\ref{DVef3}), up to first orders in $x_1$ and $x_2$, reproduces exactly the Eq.~(\ref{DVef2}), as expected.

Requiring $x_2=0$, as we have argued above, we get 
\begin{eqnarray}\label{DVef2G}
\frac{dV_\mathrm{eff}}{dB_{\mu\nu}}\Big|_{B_{\mu\nu}=\beta_{\mu\nu}} = \left[-\frac{e}{G} -\frac{e m^4}{8\pi^2} \left(\frac{1}{\epsilon'}+\frac{3}{4} \right) \left(1+\frac{x_1}2\right)^{-2}\right]b^{\mu\nu} = 0,
\end{eqnarray}
so that
\begin{equation}\label{1oG}
\frac1G = -m_R^4 \left(1+\frac{x_1}2\right)^{-2},
\end{equation}
where $m_R=Z_m^{-1/4}m$, with
\begin{equation}
\frac{1}{Z_{m}} = \frac{1}{8\pi^2} \left(\frac{1}{\epsilon'}+\frac{3}{4} \right).
\end{equation}
So, we found that $G<0$, i.e., we can write $G=-|G|$.

Now, we can rewrite the expression~(\ref{DVef2G}) as
\begin{eqnarray}
\frac{dV_\mathrm{eff}}{dB_{\mu\nu}}\Big|_{B_{\mu\nu}=\beta_{\mu\nu}} = \left[-\frac{e}{G} -em_R^4 \sum_{k=0}^\infty \frac{(-1)^k}{2^k}(k+1)x_1^k \right]b^{\mu\nu},
\end{eqnarray}
so that, by integrating it, we arrive at
\begin{eqnarray}
V_\mathrm{eff} = -\frac{1}{2G}X_1 -m_R^4 \frac{X_1}{2+X_1} +\alpha,
\end{eqnarray}
where $X_1=e^2B_{\mu\nu}B^{\mu\nu}$, $\alpha$ is an integration constant, and we have employed the power series $\sum\limits_{k=0}^\infty\frac{(-1)^k}{2^{k+1}}X_1^{k+1}=\frac{X_1}{2+X_1}$. Finally, by using (\ref{1oG}), we obtain
\begin{equation}
V_\mathrm{eff} = \frac{m_R^4}{2} X_1 \left[\left(1+\frac{x_1}2\right)^{-2} -\left(1+\frac{X_1}2\right)^{-1}\right] +\alpha.
\end{equation}
We see that this expression, first, involves arbitrary orders in fields, being non-polynomial, second, includes terms with different signs. This means that our effective potential possesses a set of minima $\langle B_{\mu\nu} \rangle$ satisfying the condition $e^2\langle B_{\mu\nu}\rangle \langle B^{\mu\nu}\rangle=b_{\mu\nu}b^{\mu\nu}$. Clearly, the most interesting situation is described by the approximation of small fields, which is more natural from the physical viewpoint, so that we can keep only some lowest orders in expansion of the effective potential in power series in $X_{1,2}$ and $x_{1,2}$. The lowest contribution to the effective potential is described by zero and first order in $x_1$ and $X_1$, and by choosing the additive constant $\alpha$ to be $\alpha=\frac{m_R^4}{4}x_1^2$, we get the simplest form of the effective potential
\begin{equation}
V_\mathrm{eff} = \frac{m_R^4}{4} (e^2B_{\mu\nu}B^{\mu\nu}-b_{\mu\nu}b^{\mu\nu})^2+\ldots,
\end{equation} 
which is the bumblebee potential for the tensorial field $B_{\mu\nu}$, and dots are for higher-order terms.

The key feature of this potential we generated is its positiveness. Therefore, it indeed possesses a set of minima where $e^2B_{\mu\nu}B^{\mu\nu}=b_{\mu\nu}b^{\mu\nu}$, so that a choice of one of these minima evidently generates a privileged spacetime direction and thus breaks the Lorentz symmetry in a spontaneous manner. So, we succeeded to generalize the methodology developed in \cite{Assuncao:2017tnz} for an antisymmetric tensor field. In principle, it is natural to expect that these calculations can be generalized as well for the finite temperature regime, and the possibility of phase transitions can be studied. We note that the main difference between tensor and pseudotensor cases consists in the fact that only in the tensor case one can obtain an exact result for the one-loop effective potential including all orders in the dynamical field. Nevertheless, in both cases, the effective potential possesses a continuous set of minima and hence allows for the spontaneous Lorentz symmetry breaking.

\section{One-loop low-energy effective action}

After we have proved that the one-loop effective potential indeed displays the minima, let us find the explicit form of the one-loop low-energy effective action, where not only potential terms are taken into account, but the second derivative terms as well.  To do this, we can rewrite the Eq.~(\ref{Seff}) as
\begin{equation}\label{Seff:2}
S_\mathrm{eff}[B] = \frac{g^2}{2} \int d^4x\, B_{\mu\nu} B^{\mu\nu} +S^{(n)}_\mathrm{eff}[B],
\end{equation}
with
\begin{equation}\label{Seffn}
S^{(n)}_\mathrm{eff}[B]= i\mathrm{Tr} \sum_{n=1}^{\infty}\frac1n \left[S(p)eB_{\mu\nu}\gamma^{[\mu}p^{\nu]}\gamma_5^q\right]^n
\end{equation}
and $S(p)=(\slashed{p}-m)^{-1}$, where we have disregarded the field independent term $-i\mathrm{Tr} \ln(\slashed{p}-m)$.

Our aim is to study the expression (\ref{Seffn}) up to the fourth order in fields, in order to obtain lower terms of the derivative and field expansion of the effective action. First,  for $n=1$ and $n=3$, evidently, $S^{(1)}_\mathrm{eff}[B]$ and $S^{(3)}_\mathrm{eff}[B]$ vanish. Then, let us focus our attention on contributions with $n=2$ and $n=4$. First, for $n=2$, we have
\begin{equation}
S^{(2)}_\mathrm{eff}[B] = \frac i2 \mathrm{Tr}\, S(p)eB_{\kappa\lambda}\gamma^{[\kappa}p^{\lambda]}\gamma_5^qS(p)eB_{\mu\nu}\gamma^{[\mu}p^{\nu]}\gamma_5^q = \frac{ie^2}{2} \int d^4x\, \Pi^{\kappa\lambda\mu\nu} B_{\kappa\lambda}B_{\mu\nu},
\end{equation}
where
\begin{equation}
\Pi^{\kappa\lambda\mu\nu} = \mathrm{tr} \int \frac{d^4p}{(2\pi)^4} S(p)\gamma^{[\kappa}p^{\lambda]}\gamma_5^qS(p-i\partial)\gamma^{[\mu}(p-i\partial)^{\nu]}\gamma_5^q.
\end{equation}
In order to calculate the above integral, we use the Feynman parametrization, so that, for $q=1$, we obtain
\begin{eqnarray}
{\cal L}^{(2)}_\mathrm{eff,1} &=& -\frac{5 e^2 m^2}{48 \pi^2 \epsilon'} B_{\mu\nu} k^2 B^{\mu\nu}+\frac{e^2m^2}{24 \pi^2 \epsilon'} B_{\mu\nu} k^{\nu} k_{\alpha} B^{\mu\alpha} +\frac{15 e^2 m^4}{48 \pi^2 \epsilon'} B_{\mu\nu} B^{\mu\nu}+\frac{e^2}{96 \pi^2 \epsilon'} B_{\mu\nu} k^4 B^{\mu\nu} \\
&+& \frac{e^2}{2880\pi^2}B_{\mu\nu}\left[(1155 m^4-430 m^2k^2 +46 k^4)-30k^4 \left(\frac{4 m^2}{k^2}-1\right)^{5/2} \csc ^{-1}\left(\frac{2 m}{\sqrt{k^2}}\right)\right]B^{\mu\nu} \nonumber\\
&-& \frac{e^2}{1440\pi^2}B_{\mu\nu}\left[(\frac{240 m^4}{k^2}-110 m^2+3 k^2)-60 m^2 \left(\frac{4 m^2}{k^2}-1\right)^{3/2} \csc ^{-1}\left(\frac{2 m}{\sqrt{k^2}}\right)\right]k^{\nu} k_{\alpha} B^{\mu\alpha}, \nonumber
\end{eqnarray}
whereas, for $q=2$, we have
\begin{eqnarray}
{\cal L}^{(2)}_\mathrm{eff,2} &=& -\frac{e^2m^2}{16\pi^2\epsilon'} B_{\mu\nu}(k^2B^{\mu\nu}-2k^\nu k_\alpha B^{\mu\alpha}) +\frac{e^2m^4}{16\pi^2\epsilon'} B_{\mu\nu} B^{\mu\nu} +\frac{e^2}{96\pi^2\epsilon'} B_{\mu\nu} k^4B^{\mu\nu}  \nonumber\\
&&+\frac{e^2}{576\pi^2} B_{\mu\nu} \left[(27m^4-42m^2k^2+8k^4)+6k^4\left(\frac{4m^2}{k^2}-1\right)^{3/2}\mathrm{csc}^{-1}\left(\frac{2m}{\sqrt{k^2}}\right)\right] B^{\mu\nu} \nonumber\\
&&+\frac{e^2}{96\pi^2} B_{\mu\nu} \left[(18m^2-k^2)-12m^2\left(\frac{4m^2}{k^2}-1\right)^{1/2}\mathrm{csc}^{-1}\left(\frac{2m}{\sqrt{k^2}}\right)\right] k^\nu k_\alpha B^{\mu\alpha},
\end{eqnarray}
with the external momentum being related with the derivative of the field through the relation $k_{\mu} = i\partial_{\mu}$, where we have taken into account that the effective action is the integral from the effective Lagrangian over the spacetime, $S^{(2)}_\mathrm{eff}=\int d^4x\,{\cal L}^{(2)}_\mathrm{eff}$.

Now, by imposing the limit of slowly varying fields, which is formally written as $\partial^2\ll m^2$ (while $m\neq0$), for $q=1$, we get
\begin{eqnarray}
{\cal L}^{(2)}_\mathrm{eff,1} &=& \frac{5e^2m^2}{48 \pi^2}  \left(\frac{1}{\epsilon'}+\frac{1}{2}\right) B_{\mu\nu}(\partial^2B^{\mu\nu}-2\partial^\nu \partial_\alpha B^{\mu\alpha}) +\frac{5 e^2 m^4 }{16 \pi ^2} \left(\frac{1}{\epsilon'}+\frac{3}{4}\right)  B_{\mu\nu} B^{\mu\nu} \nonumber\\
&& +\frac{e^2 m^2 }{6 \pi^2} \left(\frac{1}{\epsilon'}+\frac{1}{2}\right) B_{\mu\nu} \partial^\nu \partial_\alpha B^{\mu\alpha}+\frac{e^2}{96\pi^2 \epsilon'} B_{\mu\nu} \partial^4 B^{\mu\alpha}+{\cal O}\left(\frac{\partial^4}{m^4}\right),
\end{eqnarray}
and, for $q=2$,
\begin{eqnarray}
{\cal L}^{(2)}_\mathrm{eff,2} &=& \frac{e^2m^2}{16\pi^2} \left(\frac{1}{\epsilon'}+\frac12\right) B_{\mu\nu}(\partial^2B^{\mu\nu}-2\partial^\nu \partial_\alpha B^{\mu\alpha}) +\frac{e^2m^4}{16\pi^2} \left(\frac{1}{\epsilon'}+\frac34\right) B_{\mu\nu} B^{\mu\nu} \nonumber\\
&&+\frac{e^2}{96\pi^2\epsilon'} B_{\mu\nu} \partial^4B^{\mu\nu} +{\cal O}\left(\frac{\partial^2}{m^2}\right).
\end{eqnarray}
We can rewrite the above expressions as
\begin{eqnarray}
{\cal L}^{(2)}_\mathrm{eff,1} &=& \frac{1}{4Z_3} B_{\mu\nu}(\partial^2B^{\mu\nu}-2\partial^\nu \partial_\alpha B^{\mu\alpha})+\frac{e^2m_R^4}{2} B_{\mu\nu} B^{\mu\nu}+\frac{2}{5Z_3} B_{\mu\nu} \partial^\nu \partial_\alpha B^{\mu\alpha} +{\cal O}\left(\frac{\partial^2}{m^2}\right),
\end{eqnarray}
and
\begin{eqnarray}
{\cal L}^{(2)}_\mathrm{eff,2} &=& \frac{1}{4Z_3} B_{\mu\nu}(\partial^2B^{\mu\nu}-2\partial^\nu \partial_\alpha B^{\mu\alpha}) +\frac{e^2m_R^4}{2} B_{\mu\nu} B^{\mu\nu} +{\cal O}\left(\frac{\partial^2}{m^2}\right),
\end{eqnarray}
where
\begin{equation}
\frac{1}{Z_3} = \frac{5e^2m^2}{12\pi^2} \left(\frac{1}{\epsilon'}+\frac12\right).
\end{equation}
and
\begin{equation}
\frac{1}{Z_{3}} = \frac{e^2m^2}{4\pi^2} \left(\frac{1}{\epsilon'}+\frac12\right),
\end{equation}
respectively. By defining the renormalized field $B_R^{\mu\nu}=Z_3^{-1/2}B^{\mu\nu}$, as well as the renormalized coupling constant $e_R=Z_3^{1/2}e$, for $q=1$, we obtain 
\begin{eqnarray}\label{Seff25}
{\cal L}^{(2)}_\mathrm{eff,1} &=& -\frac{1}{12} H_{R\mu\nu\lambda}H_R^{\mu\nu\lambda}-\frac{2}{5} (\partial_\alpha B_R^{\mu\alpha})^2 +\frac{e_R^2m_R^4}{2} B_{R\mu\nu} B_R^{\mu\nu},
\end{eqnarray}
and, for $q=2$,
\begin{eqnarray}\label{Seff2}
{\cal L}^{(2)}_\mathrm{eff,2} &=& -\frac{1}{12} H_{R\mu\nu\lambda}H_R^{\mu\nu\lambda} +\frac{e_R^2m_R^4}{2} B_{R\mu\nu} B_R^{\mu\nu},
\end{eqnarray}
where we have disregarded the terms contributing to higher orders of the derivative expansion.

Finally, for $n=4$, we have
\begin{eqnarray}
S^{(4)}_\mathrm{eff}[B] &=& \frac i4 \mathrm{Tr}\, S(p)eB_{\alpha\beta}\gamma^{[\alpha}p^{\beta]}\gamma_5^qS(p)eB_{\gamma\delta}\gamma^{[\gamma}p^{\delta]}\gamma_5^qS(p)eB_{\kappa\lambda}\gamma^{[\kappa}p^{\lambda]}\gamma_5^qS(p)eB_{\mu\nu}\gamma^{[\mu}p^{\nu]}\gamma_5^q \nonumber\\
&=& \frac{ie^4}{4} \int d^4x\, \Pi^{\alpha\beta\gamma\delta\kappa\lambda\mu\nu} B_{\alpha\beta}B_{\gamma\delta}B_{\kappa\lambda}B_{\mu\nu},
\end{eqnarray}
where
\begin{equation}
\Pi^{\alpha\beta\gamma\delta\kappa\lambda\mu\nu} = \mathrm{tr} \int \frac{d^4p}{(2\pi)^4} S(p)\gamma^{[\alpha}p^{\beta]}\gamma_5^qS(p)\gamma^{[\gamma}p^{\delta]}\gamma_5^qS(p)\gamma^{[\kappa}p^{\lambda]}\gamma_5^qS(p)\gamma^{[\mu}p^{\nu]}\gamma_5^q +{\cal O}\left(\partial^4\right).
\end{equation}
Repeating the calculations of the integrals over momenta carried out above, we arrive at the following results for the fourth-order contribution to the effective Lagrangian:
\begin{equation}
{\cal L}^{(4)}_\mathrm{eff,1} = -\frac{35e^4m^4}{192\pi^2} \left(\frac{1}{\epsilon'}+\frac34\right)(B_{\kappa\lambda}B^{\kappa\lambda}B_{\mu\nu}B^{\mu\nu}+2B_{\kappa\lambda}B^{\lambda\mu}B_{\mu\nu}B^{\nu\kappa}),
\end{equation}
for $q=1$, and
\begin{equation}
{\cal L}^{(4)}_\mathrm{eff,2} = -\frac{e^4m^4}{64\pi^2} \left(\frac{1}{\epsilon'}+\frac34\right)(B_{\kappa\lambda}B^{\kappa\lambda}B_{\mu\nu}B^{\mu\nu}+2B_{\kappa\lambda}B^{\lambda\mu}B_{\mu\nu}B^{\nu\kappa}),
\end{equation}
for $q=2$, where we have disregarded the derivative terms, which contribute only to higher orders of the expansion. Now, by using the identity (\ref{b3}), with $b_{\mu\nu}$ replaced by $B_{\mu\nu}$, we can rewrite the above expressions as
\begin{eqnarray}\label{Seff45}
{\cal L}^{(4)}_\mathrm{eff,1} &=& -\frac{7e_R^4 m_R^4}{12} B_{R\kappa\lambda}B_R^{\kappa\lambda}B_{R\mu\nu}B_R^{\mu\nu} -\frac{7e_R^4m_R^4}{48} B_{R\kappa\lambda}\tilde{B}_R^{\kappa\lambda}B_{R\mu\nu}\tilde{B}_R^{\mu\nu},
\end{eqnarray}
for $q=1$, and
\begin{equation}\label{Seff4}
{\cal L}^{(4)}_\mathrm{eff,2} = -\frac{e_R^4m_R^4}{4} B_{R\kappa\lambda}B_R^{\kappa\lambda}B_{R\mu\nu}B_R^{\mu\nu} -\frac{e_R^4m_R^4}{16} B_{R\kappa\lambda}\tilde{B}_R^{\kappa\lambda}B_{R\mu\nu}\tilde{B}_R^{\mu\nu},
\end{equation}
for $q=2$, where we have defined 
\begin{equation}
\frac{1}{Z_{m}} = \frac{5}{8\pi^2} \left(\frac{1}{\epsilon'}+\frac{3}{4} \right)
\end{equation}
and
\begin{equation}
\frac{1}{Z_{m}} = \frac{1}{8\pi^2} \left(\frac{1}{\epsilon'}+\frac{3}{4} \right),
\end{equation}
respectively.

Therefore, using (\ref{Seff:2}), (\ref{Seff25}), and (\ref{Seff45}), we arrive at the complete expression for the  low-energy tensorial bumblebee Lagrangian, for $q=1$, i.e. in the pseudotensor case, given by
\begin{eqnarray}
{\cal L}_{B,1} &=& -\frac{1}{12} H_{R\mu\nu\lambda}H_R^{\mu\nu\lambda} +\frac{e_R^2m_R^4}{2} B_{R\mu\nu} B_R^{\mu\nu} +\frac{e_R^2}{2G}B_{R\mu\nu}B_R^{\mu\nu} -\frac{7e_R^4m_R^4}{12} B_{R\kappa\lambda}B_R^{\kappa\lambda}B_{R\mu\nu}B_R^{\mu\nu} \nonumber\\
&&-\frac{7e_R^4m_R^4}{48} B_{R\kappa\lambda}\tilde{B}_R^{\kappa\lambda}B_{R\mu\nu}\tilde{B}_R^{\mu\nu}-\frac{2}{5} (\partial_\alpha B_R^{\mu\alpha})^2,
\end{eqnarray}
and taking into account (\ref{Seff:2}), (\ref{Seff2}), and (\ref{Seff4}) as well, for $q=2$, i.e. in the tensor case, we find
\begin{eqnarray}
{\cal L}_{B,2} &=& -\frac{1}{12} H_{R\mu\nu\lambda}H_R^{\mu\nu\lambda} +\frac{e_R^2m_R^4}{2} B_{R\mu\nu} B_R^{\mu\nu} +\frac{e_R^2}{2G}B_{R\mu\nu}B_R^{\mu\nu} -\frac{e_R^4m_R^4}{4} B_{R\kappa\lambda}B_R^{\kappa\lambda}B_{R\mu\nu}B_R^{\mu\nu} \nonumber\\
&&-\frac{e_R^4m_R^4}{16} B_{R\kappa\lambda}\tilde{B}_R^{\kappa\lambda}B_{R\mu\nu}\tilde{B}_R^{\mu\nu}.
\end{eqnarray}
To simplify these expressions, we can use $\frac1G=-m_R^4(1-\frac73 b_{\mu\nu}b^{\mu\nu})$ of Eq.~(\ref{DVef25}), for $q=1$, and $\frac1G=-m_R^4(1-b_{\mu\nu}b^{\mu\nu})$ of Eq.~(\ref{DVef2}) (as well as of (\ref{1oG}) taken up to first order in $x_1$), for $q=2$, so that we obtain
\begin{eqnarray}
{\cal L}_{B,1} &=& -\frac{1}{12} H_{R\mu\nu\lambda}H_R^{\mu\nu\lambda} -\frac{7 m_R^4}{12} (e_R^2B_{R\mu\nu}B_R^{\mu\nu}-b_{\mu\nu}b^{\mu\nu})^2 -\frac{2}{5} (\partial_\alpha B_R^{\mu\alpha})^2 -\frac{7e_R^4m_R^4}{48} (B_{R\mu\nu}\tilde{B}_R^{\mu\nu})^2, \nonumber\\
\end{eqnarray}
and
\begin{eqnarray}
{\cal L}_{B,2} &=& -\frac{1}{12} H_{R\mu\nu\lambda}H_R^{\mu\nu\lambda} -\frac{m_R^4}{4} (e_R^2B_{R\mu\nu}B_R^{\mu\nu}-b_{\mu\nu}b^{\mu\nu})^2 -\frac{e_R^4m_R^4}{16} (B_{R\mu\nu}\tilde{B}_R^{\mu\nu})^2,
\end{eqnarray}
where we have added the constants $-\frac{7m_R^4}{12}x_1^2$ and $-\frac{m_R^4}{4}x_1^2$, respectively. 

Thus, we conclude that we have succeeded to generate the low-energy effective actions for the tensor bumblebee field, which includes the usual kinetic term for the second-rank antisymmetric tensor field, the positively defined potential (displaying the set of minima $x_1=2(\sqrt{|G|m_R^4}-1)$, for $q=2$, according to Eq.~(\ref{DVef2G})), allowing for spontaneous Lorentz symmetry breaking, and another one with the trivial minimum $x_2=0$. We note that both currents, corresponding both to $q=1$ and $q=2$ (pseudotensor and tensor cases), display rather similar dynamical impacts allowing to achieve a set of minima, allowing for spontaneous Lorentz symmetry breaking, in both cases. We also observe that other, derivative-free forms of the second-rank tensor current, such as $J_{\mu\nu}=i\bar{\psi}\sigma_{\mu\nu}\gamma_5^q\psi$, do not possess this feature, i.e., the one-loop effective potentials generated with their use do not display minima (see Eq.~(\ref{DVef2J2})), hence, the spontaneous Lorentz symmetry breaking cannot occur for these couplings.

\section{Summary}

Now, let us discuss our results. Within this paper, we have successfully generalized the mechanism of the dynamical Lorentz symmetry breaking for theories of the second-rank antisymmetric tensor and pseudotensor fields, i.e., for the first time, we have generated the tensor bumblebee action as a quantum correction, while, earlier, only the vector bumblebee model was studied within the perturbative methodology. Therefore, it is natural to expect that our results can open the way for further studies of spontaneous Lorentz symmetry breaking for generic dynamical tensor fields. The approach we have used continues the line of our earlier paper \cite{Assuncao:2017tnz} and guarantees that our effective potential indeed possesses minima, which justifies the consistency of our results. We note that, actually, our paper represents itself as one of the first studies of quantum aspects of Lorentz symmetry breaking for the higher-rank tensor field models. Especially, it must be emphasized that the one-loop effective potential we have obtained displays a continuous set of minima both in tensor and pseudotensor cases, hence, in both these cases the spontaneous Lorentz symmetry breaking can occur. It is important to note that unlike the previous papers \cite{Colatto:2003he,Hernaski:2016dyk,Aashish:2018aqn,Aashish2}, in our paper, the constant antisymmetric second-rank pseudotensor  has been introduced for the first time, which opens the way for constructing new Lorentz-breaking terms involving such a pseudotensor.

It is natural to expect that our methodology can be applied to more sophisticated tensor models. First, it is important to note that apparently our results can be useful within the string context. To present a possible relation of our results to string theory, it is worth to mention that, from one side, the spontaneous Lorentz symmetry breaking has been originally introduced namely within the string context \cite{Kostelecky:1989jp}, and from another side, the antisymmetric tensor field arises within the low-energy limit of the string theory \cite{Callan}. Therefore, actually our paper explicitly demonstrates the essence of the mechanism proposed in \cite{Kostelecky:1989jp}.  Thus, it is natural to expect that our results can be applied for detailed studies of various low-energy consequences of the string theory and of different higher-rank tensor models. Another continuation of the present work could consist in introducing the finite temperature, with the subsequent study of the possibility of phase transitions, generalizing the results of \cite{Assuncao:2017tnz} for a tensor field case, as well as in introducing of a curved background extending thus a study carried out in \cite{Altschul:2009ae} for a level of quantum corrections. We expect to carry out these generalizations in forthcoming papers. 

{\bf Acknowledgements.} This work was partially supported by Conselho
Nacional de Desenvolvimento Cient\'{\i}fico e Tecnol\'{o}gico (CNPq). The work by A. Yu. P. has been supported by the
CNPq project No. 303783/2015-0.

\end{document}